\documentclass[twocolumn,showpacs,letterpaper]{revtex4}

\usepackage{amsmath,amssymb}
\usepackage[pdftex]{graphicx}

\newcommand{\bra}[1]{\langle #1|}
\newcommand{\ket}[1]{|#1\rangle}

\begin{document}

\title{Dimer-dimer collisions at finite energies in two-component Fermi gases}

\author{J. P. D'Incao}
\author{Seth T. Rittenhouse}
\author{N. P. Mehta}
\author{Chris H. Greene}
\affiliation{Department of Physics and JILA, University of Colorado,  
Boulder, Colorado 80309-0440, USA}

\begin{abstract}
We introduce a major theoretical generalization of existing techniques for handling the three-body problem 
that accurately describes the interactions among four fermionic atoms. Application to a two-component 
Fermi gas accurately determines dimer-dimer scattering parameters at finite energies and can give deeper 
insight into the corresponding many-body phenomena. To account for finite temperature effects, 
we calculate the energy-dependent  {\em complex} dimer-dimer scattering length, 
which includes contributions from elastic and inelastic collisions. 
Our results indicate that strong finite-energy effects 
{and dimer dissociation}
are crucial for understanding 
the physics in the strongly interacting regime for typical experimental conditions. 
While our results for dimer-dimer relaxation are consistent with experiment, 
they confirm only partially a previously published theoretical result.
\end{abstract} 
\pacs{31.15.xj,34.50.-s,34.50Cx,67.85.-d}

\maketitle 
 
The physics of strongly interacting fermionic systems is of fundamental importance in many areas of physics
encompassing condensed matter physics, nuclear physics, particle physics and astrophysics.
The last few years have seen extensive theoretical and experimental efforts devoted to the field of
ultracold atomic Fermi gases. The ability to control interatomic interactions through magnetically 
tunable Feshbach resonances has opened up broad vistas of experimentally accessible phenomena, providing a quantum playground for studying the strongly interacting regime.
For instance, near a Feshbach resonance between two dissimilar fermions,
the $s$-wave scattering length $a$ can assume positive and negative values, allowing for the systematic exploration
of Bose-Einstein condensation (BEC) and the Bardeen-Cooper-Schrieffer (BCS) crossover regime, in which 
bosonic ($a>0$) and fermionic ($a<0$) types of superfluidity connect smoothly \cite{CrossOverTheory,CrossOverExp}.  
In this broad context, few-body correlations \cite{Strinati,Petrov} play an important role in describing the dynamics of such systems.
On the BEC side of the resonance ($a>0$), dissimilar fermions pair-up into weakly-bound bosonic dimers, and the 
zero (collision) energy dimer-dimer scattering length, $a_{dd}(0)$,
determines various experimental observables such as the molecular gas collective modes, the internal energy, and even the macroscopic 
spatial extent of the confined cloud \cite{CrossOverTheory,CrossOverExp}. 
Although a better description of the many-body behavior has emerged 
through the inclusion of few-body correlations, most of the current understanding of crossover physics relies 
on zero-energy theories, and very little is known about finite energy effects in this regime
 (see Ref.~\cite{FiniteTemp} and references therein).

In this Letter we demonstrate important finite energy effects which can potentially impact the physics of a 
finite temperature ultracold Fermi gas in the crossover regime.
Our results show deviations from zero-energy dimer-dimer collisions
and indicate that, at experimentally relevant temperatures and scattering lengths, molecular dissociation might play an important role.
The crossover regime can be viewed as a long-lived atom-molecule mixture, where dimers are dynamically converted to atoms and vice-versa.
In order to account for finite temperature effects, we calculate the energy dependent  
{\em complex} dimer-dimer scattering length, $a_{dd}(E_{col})$, where $E_{col}$ is the collision energy.  
The real and imaginary parts of $a_{dd}$ correspond, respectively, to contributions from elastic and 
inelastic (dissociative) collisions \cite{ComplexScatLen}, both of which should be considered to
properly model the Fermi gas at realistic temperatures. 
In the zero-energy limit we reproduce the well known prediction $a_{dd}(0)\approx0.6a$ \cite{Petrov}.
However, when the dimer binding energy, $E_{b}=\hbar^2/(2\mu_{\rm 2b} a^2)$ (where $\mu_{\rm 2b}$ is the 
two-body reduced mass) is comparable to the gas temperature $T$, finite energy effects and molecular dissociation become important, 
defining a critical scattering length $a_{c}=\hbar/(2\mu_{\rm 2b} k_{B}T)^{1/2}$, where $k_{B}$ is Boltzmann's constant,
beyond which  an atom-molecule mixture should prevail.

We also study dimer-dimer relaxation, in which two weakly-bound dimers
collide and make an inelastic transition to a lower energy state. In such a process, the kinetic energy released is
enough for the collision partners to escape from typical traps. 
Ref.~\cite{Petrov} predicted that, near a Feshbach resonance, dimer-dimer relaxation is suppressed as $a^{-2.55}$, 
explaining the long lifetimes observed in several experiments \cite{CrossOverExp,LongLivedMol}.
Here we also verify this suppression, although with an $a$ dependence that is not described as a simple 
power-law scaling as originally predicted \cite{Petrov}.
While the $a^{-2.55}$ scaling law has already been tested (Regal {\em et al.} \cite{LongLivedMol} 
found $a^{-2.3\pm0.4}$ and Bourdel {\em et al.} \cite{CrossOverExp} $a^{-2.0\pm0.8}$), 
our calculations demonstrate that 
finite range corrections can explain the apparent experimental
scaling law behavior, despite deviations from that power-law for larger $a$.

We solve the four-body Schr\"odinger equation in the hyperspherical adiabatic representation, 
which offers a simple yet quantitative picture. A finite range model is assumed for the interatomic interaction, 
and a physically-motivated variational basis set is adopted to solve the hyperangular equations \cite{FourBodyUs}. 
While several hyperangular parameterizations exist, we find that the best choice is the ``democratic'' hyperspherical 
coordinates \cite{HypCoord} in which all possible fragmentation channels are treated on an equal footing, 
which describes elastic and inelastic processes in an unified picture.

In the adiabatic hyperspherical representation, the collective motion of the four fermions is described in terms of the 
hyperradius $R$, characterizing the overall size of the system. 
The interparticle relative motion is described 
by the hyperangles $\Omega\equiv\{\theta_{1},\theta_{2},\phi_{1},\phi_{2},\phi_{3}\}$, 
and the set of Euler angles $\{\alpha$, $\beta$, $\gamma\}$ specifying the orientation of the body-fixed frame 
\cite{HypCoord}. 
$\theta_{1}$ and $\theta_{2}$ parameterize the moments of inertia while $\phi_{1}$, $\phi_{2}$ and 
$\phi_{3}$ parameterize internal configurations \cite{HypCoord}.
Integrating out the hyperangular degrees of freedom, the Schr\"odinger 
equation reduces to a system of coupled ordinary differential equations, given in atomic units (used throughout this Letter) by:
\begin{eqnarray}
\left[-\frac{1}{2\mu}\frac{d^2}{dR^2}-E\right]F_{\nu}(R)+\sum_{\nu'} W_{\nu\nu'}(R) F_{\nu'}(R)=0,
\label{SchrEq}
\end{eqnarray}
\noindent
where $\mu=m/4^{\frac{1}{3}}$ is the four-body reduced mass ($m$ being the atomic mass), $E$ is the total energy, 
$F_{\nu}$ is the hyperradial wavefunction, and $\nu$ represents all quantum numbers 
needed to label each channel. Scattering observables can then be extracted by solving Eq.~(\ref{SchrEq}),
where the nonadiabatic couplings $W_{\nu\ne\nu'}$ drive inelastic transitions between channels described
by the effective potentials $W_{\nu\nu}$.

\begin{figure}[htbp]
\includegraphics[width=3.in,angle=0,clip=true]{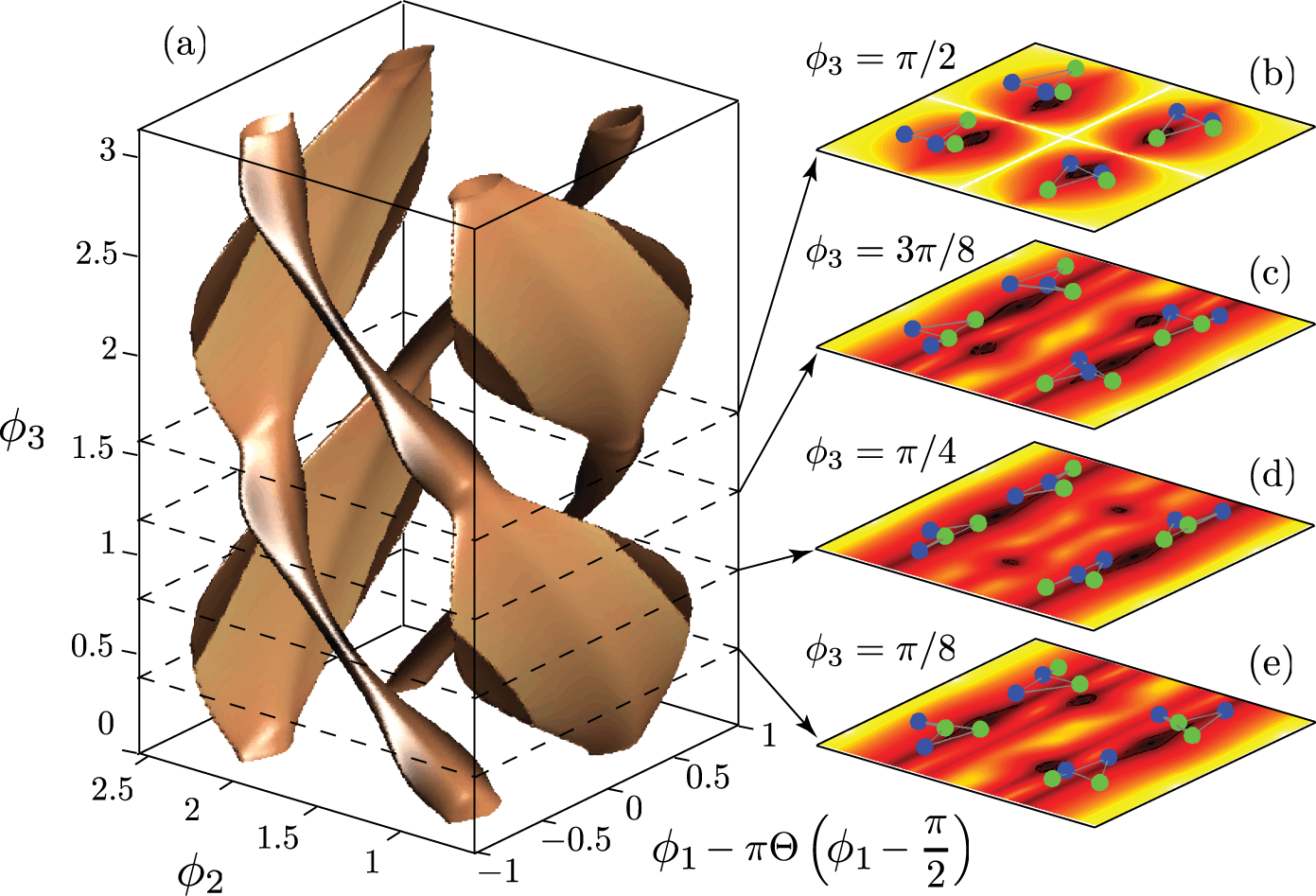}
\caption{(color online). Analysis of the probability density integrated over the hyperangles $\theta_{1}$ and $\theta_{2}$ 
at $R=0.41~a$ is shown.  
(a) An isosurface of the probability density at $|\Phi(R;\Omega)|^2=0.1|\Phi(R;\Omega)|^2_{\rm max}$
for the dimer-dimer channel is shown.
The darker (lighter) colors correspond to a more (less) linear configuration for the four-particle system. $\Theta(x)$ is the 
unit-step function. 
(b)-(e) {show density plots
for fixed values of $\phi_{3}$. The darker regions represent higher probabilities for which planar configurations 
are shown to illustrate the most probable four-body geometry at selected points.}}
\label{ChannelFig}
\end{figure}

In the hyperspherical representation, the major reduction to Eq.~(\ref{SchrEq}) is accomplished by finding eigenfunctions
of the (fixed $R$) adiabatic Hamiltonian,
\begin{equation}
\hat{H}_{\rm ad}(R,\Omega)=
\frac{\hat{\Lambda}^2(\Omega)+12}{2\mu R^2}+\hat{V}(R,\Omega).
\label{Had}
\end{equation}
\noindent
In the above equation, $\hat\Lambda$ is the grand angular momentum operator \cite{HypCoord} and $\hat{V}$ 
includes all two-body interactions \cite{Euler}.
For simplicity, we neglect the interaction between identical 
fermions and assume the one between dissimilar fermions to be $v(r)=D{\rm sech}^2(r/r_{0})$, where $r$ is 
the interatomic distance and $D$ is tuned to produce the desired $a$. 
We choose the atomic mass $m$ and effective range $r_{0}=181$~a.u. \cite{Flambaum}
to be those of $^{40}$K.
The eigenvalues and eigenfunctions of $\hat{H}_{\rm ad}$, namely the
hyperspherical potentials $U_{\nu}(R)$ and channel functions $\Phi_{\nu}(R;\Omega)$, determine
the effective potentials and nonadiabatic couplings in Eq.~(\ref{SchrEq}): 
{$W_{\nu\nu}=U_{\nu}-Q_{\nu\nu}/{2\mu}$ and 
$W_{\nu\nu'}=-\left[P_{\nu\nu'}{d}/{dR}+Q_{\nu\nu'}\right]/{2\mu}$,
where $P_{\nu\nu'}=\bra{\Phi_{\nu}}{d}/{dR}\ket{\Phi_{\nu'}}$ and 
$Q_{\nu\nu'}=\bra{\Phi_{\nu}}{d^2}/{dR^2}\ket{\Phi_{\nu'}}$.}
We find $\Phi_{\nu}(R;\Omega)$ variationally by expanding in exact eigenfunctions of Eq.~(\ref{Had})
at large {\em and} small $R$ \cite{FourBodyUs}. 
At ultracold energies the convergence of the scattering observables with respect to the number 
of basis functions is surprisingly fast \cite{FourBodyUs}.

\begin{figure}[htbp]
\includegraphics[width=3.3in,angle=0,clip=true]{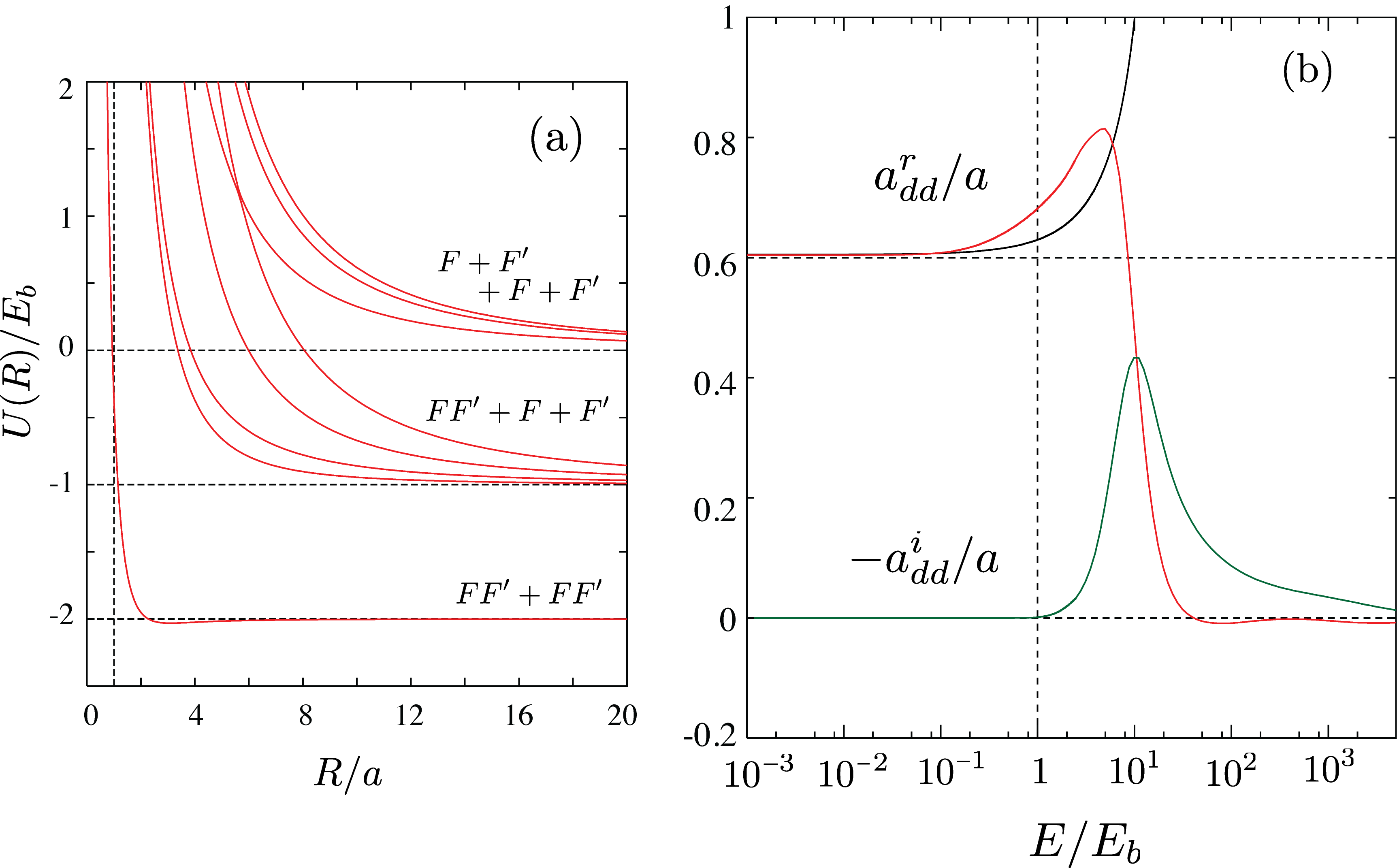}
\caption{(color online). 
(a) Several four-fermion hyperspherical potentials attached to all relevant breakup thresholds are shown. 
(b) The energy-dependent elastic (red) and inelastic (green)  parts of $a_{dd}$ [Eq.~(\ref{addE})] are shown. 
For energies $E\ll E_b$  we find
$a^{r}_{dd}=0.605(5)a$ \cite{Petrov},  while for $E_{col}\approx E_{b}$  
finite energy corrections strongly affect $a_{dd}$. Solid black line: $a_{dd}$ obtained from 
the effective range expansion~\cite{Javier}.}
\label{PotentialsAdd}
\end{figure}

\begin{figure*}
\includegraphics[width=6.2in,angle=0,clip=true]{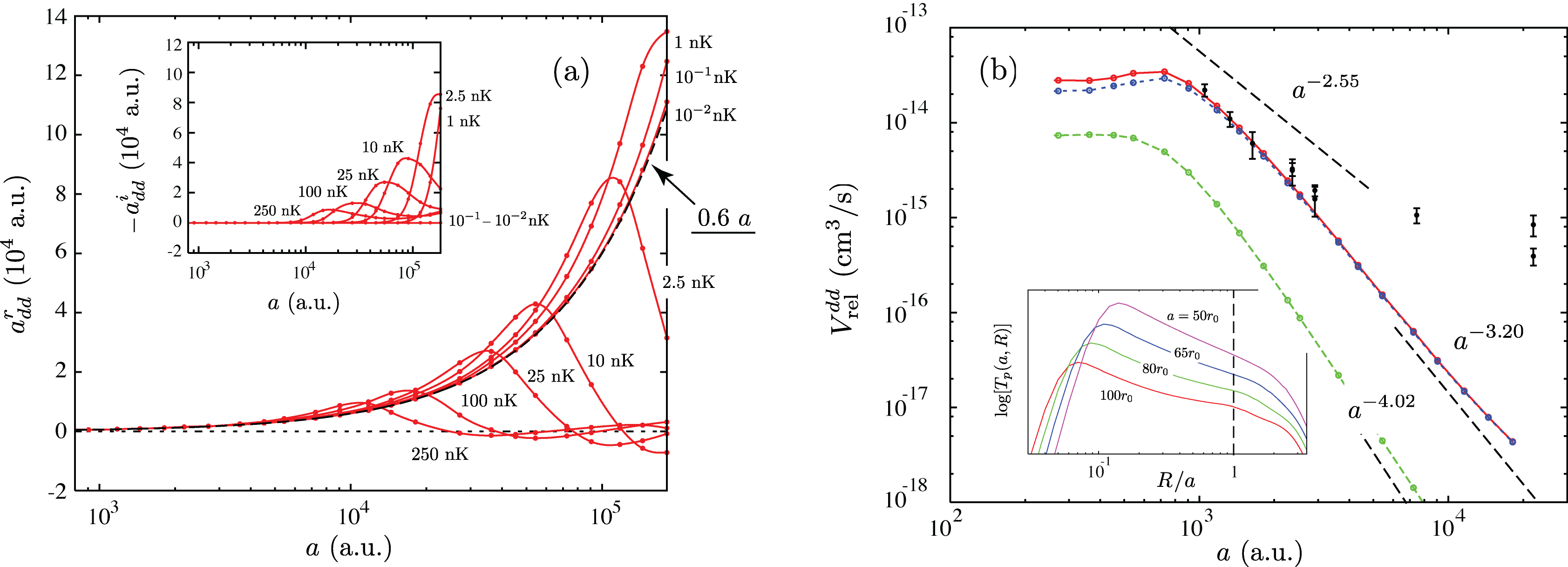}
\caption{(color online). 
(a) The scattering length dependence of $a_{dd}^r$ and $a_{dd}^{i}$ (inset)
at fixed collision energies is shown.
At {\em any} non-zero collision energy $a_{dd}$ deviates from the zero-energy prediction (black dashed line) as $a\rightarrow\infty$
and strong contributions from molecular dissociation, $a_{dd}^{i}$, occur.
(b) The dimer-dimer relaxation rate, $V_{\rm rel}^{dd}$, is shown versus $a$. The solid line is the total $V_{\rm rel}^{dd}$ and the dashed lines
are contributions from different decay pathways (see text). For intermediate $a$, 
we reproduce both experimental data \cite{LongLivedMol} (filled circles) and the $a^{-2.55}$ scaling 
law \cite{Petrov} while deviating from that for larger $a$.
Inset: we show $T_{p}$ as a function of $R$ for $a=100r_0$, $80r_{0}$, $65r_{0}$, and $50r_0$ (red, green, blue, and magenta respectively).}
\label{AddFig}
\end{figure*}

However, including higher order correlations that describe dimer-atom-atom and four free atom
configurations is crucial for {\em accurately} describing scattering processes at {\em any} collision energy.
We find that for $R\lesssim a$ the strongest 
contribution to the probability density of the dimer-dimer channel function 
comes from dimer-atom-atom like configurations. 
Figure~\ref{ChannelFig} shows a graphical representation of this channel function {in terms of the internal configuration angles $\phi_1$,
$\phi_2$ and $\phi_3$}.
The four ``cobra''-like surfaces explicitly illustrate the four-fold symmetry ($S_{2}\otimes S_{2}$) 
of the fermionic problem. The ``spines'' of the cobras correspond to the interaction valleys where two dissimilar 
fermions are in close proximity while the ``hoods'' loosely represent the larger phase-space explored by 
dimer-atom-atom like configurations.

Figure \ref{PotentialsAdd}(a) shows the hyperspherical potentials for $a=125r_{0}$
showing the full energy landscape with the four-body thresholds for dimer-dimer ($FF'+FF'$), dimer-atom-atom 
($FF'+F+F'$), and four atom ($F+F'+F+F'$) collisions. 
Notice that the four-body potential associated with dimer-dimer collisions is repulsive for $R<a$,
indicating that zero-energy dimer-dimer elastic scattering must be qualitatively similar to scattering by a 
hard-sphere of radius $a$, i.e., $a_{dd}(0)\propto a$.
Although a clear and qualitative picture emerges from 
the four-body potentials alone, we in practice
extract scattering observables from coupled-channel solutions to Eq.~(\ref{SchrEq}). 
We define the energy dependent dimer-dimer scattering
length, $a_{dd}(E_{col})$, in terms of the complex phase-shift obtained from the corresponding $S$-matrix element 
[$S_{dd,dd}=\exp(2i\delta_{dd})$], 
\begin{equation}
a_{dd}(E_{col})=-\frac{\tan\delta_{dd}}{k_{dd}}=a_{dd}^{r}(E_{col})+i~a_{dd}^{i}(E_{col}).\label{addE}
\end{equation}
\noindent
Here, $k_{dd}^2=2mE_{col}$, $E_{col}=E+2E_{b}$ is the collision energy and $a_{dd}^{r}$ and $a_{dd}^{i}<0$ are the real and imaginary parts of $a_{dd}$, 
representing elastic and inelastic contributions \cite{ComplexScatLen}.

Figure~\ref{PotentialsAdd}(b) shows $a_{dd}^{r}$ and $a_{dd}^{i}$ for $a=125r_0$. 
For energies $E_{col}\ll E_{b}$ we find that $a_{dd}(0)=
0.605(5)a$ in agreement with Refs.~\cite{Petrov,Javier} while for $E_{col}\lesssim E_{b}$, 
although molecular dissociation is still not allowed, i.e. 
$a^{i}_{dd}=0$, we obtain strong corrections to the zero-energy result. 
At these energies, an effective range 
expansion, $a_{dd}^{-1}(E_{col})=a_{dd}^{-1}(0)-\frac{1}{2}r_{dd}k_{dd}^2$ where $r_{dd}=0.13a$ \cite{Javier}, 
is accurate over a small range, but quickly fails to reproduce our results
[see black solid line in Fig.~\ref{PotentialsAdd}(b)].
For $E_{col}\gtrsim E_{b}$, the channels for molecular dissociation become open leading to strong inelastic contributions
to $a_{dd}(E_{col})$, as parametrized by $a_{dd}^{i}$. Our results indicate that both $a_{dd}^{r}$ and $a_{dd}^{i}$ are
universal functions of energy and scattering length, i.e., insensitive to the
details of the short-range physics, which should extend up to $E_{col}\ll 1/mr_{0}^2$ in the absence of deeply bound states.
Due to the small number of basis functions used in these calculations, our results for $E_{b}\ll E_{col}\ll 1/(m r_{0}^2)$ 
are not fully converged, but we expect their qualitative behavior, i.e., the sharp decease in $a_{dd}(E_{col})$,
to persist.

Figure~\ref{AddFig}(a) demonstrates that when approaching the Feshbach resonance ($a$$\rightarrow$$\infty$)
at {\em any} finite collision energy, molecular dissociation becomes increasingly more important
and $a_{dd}^{r}$  substantially deviates from the zero-energy predictions
[black dashed line and inset in Fig.~\ref{AddFig}(a)]. As $a\rightarrow\infty$, 
$E_{b}\propto1/a^2$ becomes extremely small and such finite energy effects [see Fig.~\ref{PotentialsAdd}(b)] are relevant even at ultracold energies.
Therefore, the molecular binding energy $E_{b}$, or equivalently $a_{c}=1/\sqrt{2\mu_{\rm 2b}T}$, defines
the range beyond which (i.e., $a>_{a_{c}}$) deviations from the zero-energy predictions can be observed. Perhaps more importantly,
it specifies a regime beyond which molecular dissociation can lead to a long-lived atom-molecule mixture
\cite{AtomMoleculeMixture,AtomMoleculeMixtureAdd},
where dimers are continuously converted to atoms and vice-versa, i.e.
$FF'+FF'\leftrightarrow FF'+F+F'$.
Further, this indicates that the underlying physics of the strongly interacting regime may 
fundamentally depend on temperature.
Values for $a_{c}$ at 100~nK are 7000~a.u. for $^{40}$K 
and 17000~a.u. for $^{6}$Li, and therefore the finite energy effects above can become
experimentally relevant \cite{CrossOverExp}.

We also study vibrational relaxation due to dimer-dimer collisions. We verify the suppression of 
the relaxation rate as $a\rightarrow\infty$, however, with a different $a$ dependence than $a^{-2.55}$ predicted in Ref.~\cite{Petrov}. 
In Ref.~\cite{Petrov} it was assumed that the main decay pathway for relaxations is a purely three-body process 
and requires only three atoms to be enclosed at short distances. Therefore, it {\em neglects} the effects of the 
interaction with the fourth atom. Here, however, we analyze such effects and find that it strongly influences the suppression
of relaxation. In our calculations we express the inelastic transitions probability $T_{p}(a,R)$ in terms of the probability 
of having three atoms at short distances as a function of the distance $\approx R$ of the fourth atom from the collision center \cite{EPAPS}.
We calculate $T_{p}$ from our fully coupled-channel solutions and effective potentials [see Figs.~\ref{ChannelFig} and 
\ref{PotentialsAdd}(a)] and our results are shown in the inset of Fig.~\ref{AddFig}(b).

In our model the relaxation rate is simply proportional to the transition probability $T_{p}(a,R)$. 
It is interesting note that our formulation allows for the analysis of different decay pathways. 
For instance, at short distances, $R\approx r_{0}$, $T_{p}$ describes inelastic transitions
in which {\em all} four atoms are involved in the collision process.
At large distances, $R/a>>1$, $T_{p}$ describes the decay pathway where {\em only} three atoms participate in the collision, 
akin to the process studied in Ref.~\cite{Petrov}. We note, however, that for values of $R$ up to 
$R\approx5$ the scaling law for relaxation depends strongly on $R/a$ and greatly deviate from the $a^{-2.55}$ scaling.
In order to take into account inelastic processes for all values of $R$ we define an effective transition probability by integrating
$T_{p}(a,R)$ over $R$ \cite{EPAPS}. Our results for the relaxation rate, $V_{\rm rel}^{dd}$, are shown in Fig.~\ref{AddFig}(b) where the red solid line 
is obtained by integrating $T_{p}$ from $R=2r_{0}$ up to $10a$ \cite{MyComment} giving an apparent scaling law of $a^{-3.20\pm0.05}$. 
The dashed lines are obtained from integrating $T_{p}$ from $R=2r_{0}$ to $5r_{0}$ and from $R=5r_{0}$ to $10a$, which yields 
 scaling laws of $a^{-4.02}$ and $a^{-3.20\pm0.05}$, respectively, ``separating'' the contributions from the decay pathways in which 
four and three atoms participate in the collision process. 
The amplitudes for each of these contributions, however, are disconnected as they depend on the details of the 
four- and three-body short-range physics. 
In contrast, the amplitudes for the $a^{-3.20}$ and $a^{-2.55}$ processes are governed by the
same three-body physics. As a result, the fact that we don't observe the $a^{-2.55}$ scaling implies that it is 
not important for the range of $a$ used here.
The amplitude for the process which leads to the $a^{-2.55}$ scaling is exponentially suppressed
owing to the unfavorable overlap of the dimers' wavefunction [see inset of Fig.~\ref{AddFig}(b)].
In fact, for our largest values of $a$, it is already apparent that in the very large $a$ limit the rate deviates from
$a^{-3.20}$, however, to a behavior different than $a^{-2.55}$ \cite{EPAPS}.

Figure~\ref{AddFig}(b) rescales our results for $V_{\rm rel}^{dd}$ by an overall constant 
chosen to fit the experimental data for $^{40}$K at a temperature of $70$~nK (Regal {\em et al.} \cite{LongLivedMol}).
We note, however, that between $a=1000$ and $3000$~a.u. \cite{Comment}, our results agree with both the experimental data
{\em and} the $a^{-2.55}$ scaling law, approaching our predicted scaling law $a^{-3.20}$ only for larger values of $a$. 
This change in behavior of $V_{\rm rel}^{dd}$ originates in the
finite range of our model, which represents physics beyond the zero-range model of Ref.~\cite{Petrov} where the $a^{-2.55}$ 
scaling applies for all $a$. 

In summary, we have calculated the energy dependent {\em complex} dimer-dimer scattering length, $a_{dd}(E_{col})$,
by solving the four-body Schr\"odinger equation in the adiabatic hyperspherical representation. 
Our results demonstrate that for experimentally relevant temperatures and scattering lengths the elastic and 
inelastic contributions of $a_{dd}$ are equally important. 
We show that molecular dissociation
plays an important role and suggest that the many-body behavior in the strongly interacting regime might be
significantly altered at finite temperature. 
Our results also demonstrate a stronger suppression for dimer-dimer relaxation, compared to that obtained 
in Ref.~\cite{Petrov}, while remaining consistent with experimental data.

The authors would like to acknowledge D. S. Jin's group for providing their experimental data, J. von Stecher and D. S. Petrov for fruitful discussions,
and the W. M. Keck Foundation for providing computational resources. 
This work was supported by the National Science Foundation.


\begin{thebibliography}{99}

\bibitem{CrossOverTheory}
D. M. Eagles, Phys. Rev. {\bf 186}, 456 (1969);
A. J. Leggett, J. Phys. C {\bf 41}, 7 (1980);
M. Holland {\em et al.}, Phys. Rev. Lett. {\bf 87}, 120406 (2001);
E. Timmermans {\em et al.}, Phys. Lett. A {\bf 285}, 228 (2001);
Y. Ohashi and A. Griffin, Phys. Rev. Lett. {\bf 89}, 130402 (2002).

\bibitem{CrossOverExp} 
C. Chin {\em et al.}, Science {\bf 305}, 1128 (2004); 
T. Bourdel {\em et al.}, Phys. Rev. Lett. {\bf 93}, 050401 (2004);
M. Bartenstein {\em et al.}, {\em ibid.} {\bf 92}, 120401 (2004);
C. A. Regal, M. Greiner, and D. S. Jin, {\em ibid.} {\bf 92}, 040403 (2004);
M. W. Zwierlein {\em et al.}, {\em ibid.} {\bf 92}, 120403 (2004);
G. B. Partridge {\em et al.}, {\em ibid.}  {\bf 95} 020404 (2005); 
M. Bartenstein et al., {\em ibid.} {\bf 92}, 203201 (2004);
J. Kinast {\em et al.}, {\em ibid.} {\bf 92}, 150402 (2004).
M. Greiner, C. A. Regal, and D. S. Jin, {\em ibid.} {\bf 94}, 070403 (2005);
M. W. Zwierlein, {\em et al.}, Nature (London) {\bf 435}, 1047 (2005).

\bibitem{Strinati} P. Pieri and G. C. Strinati, Phys. Rev. B {\bf 61}, 15370 (2000).

\bibitem{Petrov} D. S. Petrov, C. Salomon, and G. V. Shlyapnikov,
  Phys. Rev. Lett. {\bf 93}, 090404 (2004); Phys. Rev. A {\bf 71}, 012708 (2005).

\bibitem{FiniteTemp} M. J. Wright {\em et al.}, Phys. Rev. Lett. {\bf 99}, 150403 (2007).

\bibitem{ComplexScatLen} J. L. Bohn and P. S. Julienne, Phys. Rev. A {\bf 56}, 1486 (1997); R. C. Forrey {\em et al.}, 
{\em ibid.} {\bf 58}, R2645 (1998).

\bibitem{LongLivedMol}
J. Cubizolles {\em et al.}, Phys. Rev. Lett. {\bf 91},   240401 (2003);
K. E. Strecker, G. B. Partridge, and R. G. Hulet, {\em ibid.} {\bf 91} 080406 (2003);
C. A. Regal, M. Greiner, and D. S. Jin,  {\em ibid.} {\bf 92}, 083201 (2004).

\bibitem{FourBodyUs} Nirav P. Mehta, Seth T. Rittenhouse, Jos\'e P. D'Incao, Chris H. Greene,
arXiv:0706.1296.

\bibitem{HypCoord} V. Aquilanti and S. Cavalli. J. Chem. Soc. : Faraday Transactions {\bf 95}, 801 (1997);
A. Kuppermann. J. Phys. Chem. A  {\bf 101}, 6368 (1997).

\bibitem{Euler} We omit all reference to the Euler angles because this study treats only the dominant 
partial wave $J^{\pi}=0^+$, where $J$ is the total orbital angular momentum and
$\pi$ is the parity quantum number.

\bibitem{Flambaum} V. V. Flambaum, G. F. Gribakin, and C. Harabati, Phys. Rev. A {\bf 59}, 1998 (1999). 
The value for $r_{0}$ is obtained from the expression 
$r_{0}=2^{1/2}3^{-1}\Gamma(\frac{1}{4})\Gamma(\frac{3}{4})^{-1}(mC_{6})^{1/4}$, Eq,~(24),
at resonance.

\bibitem{Javier} J. von Stecher, Chris H. Greene, and D. Blume, Phys. Rev. A {\bf 76}, 053613 (2007);
 {\em ibid.} {\bf 77}, 043619 (2008). $a_{dd}$ in these works is calculated to an accuracy similar to our
present value.

\bibitem{AtomMoleculeMixture} S. Jochim {\em et al.}, Phys. Rev. Lett. {\bf 91}, 240402 (2003).

\bibitem{AtomMoleculeMixtureAdd} Y. Shin, A. Schirotzek, C. H. Schunck, and W. Ketterle, Phys. Rev. Lett. {\bf 101}, 070404 (2008).

\bibitem{EPAPS} See Supplemental Material attached to this file for a more detailed description
of our model for dimer-dimer relaxation.

\bibitem{MyComment} We have found that integrating $T_{p}$ from $R=2r_{0}$ up to $10a$
is enough to ensure that contributions for $R<2r_{0}$ and $R>10a$ are neglegible.

\bibitem{Comment} The experimental data for $a>3000$~a.u. were taken
in the regime where the molecular size is expected to be larger than the interparticle spacing, which prohibits a proper 
comparisson to our results. 


\end{thebibliography}
\end{document}


\title{Supplemental Material: ``Dimer-dimer collisions at finite energies in two-component Fermi gases''}

\author{J. P. D'Incao}
\author{Seth T. Rittenhouse}
\author{N. P. Mehta}
\author{Chris H. Greene}
\affiliation{Department of Physics and JILA, University of Colorado,  
Boulder, Colorado 80309-0440, USA}

\maketitle 
 
In this supplemental material we provide additional details of our model for dimer-dimer relaxation
in terms of the transition probability obtained from our numerical calculations \cite{Us}. 

\begin{figure}[htbp]
\includegraphics[width=3.0in,angle=0,clip=true]{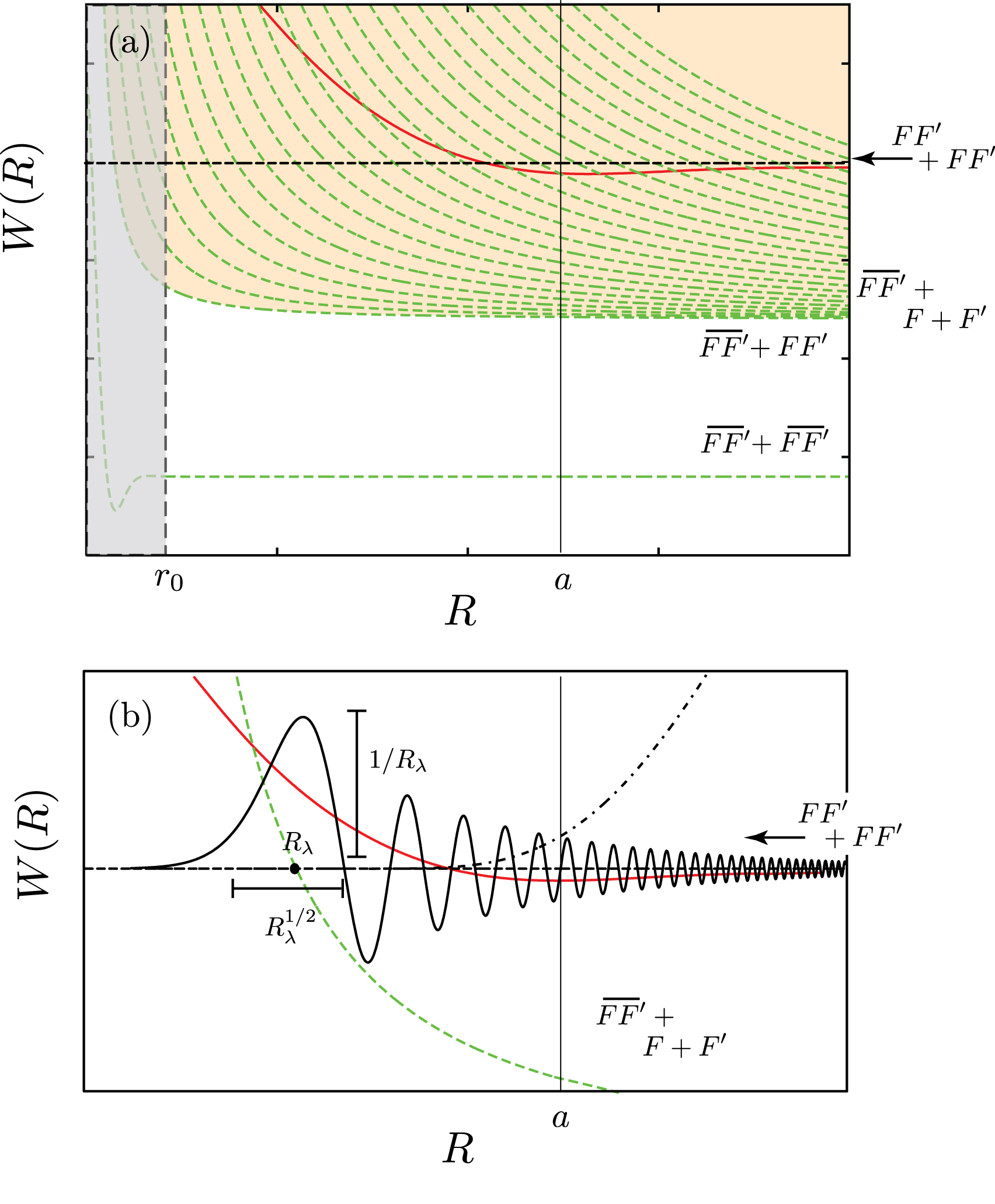}
\caption{(color online). (a) A schematic representation of the dimer-dimer relaxation process is shown. The red-solid solid curve represents the 
effective potential for the initial dimer-dimer channel, $FF'+FF'$, and green-dashed curves represent the possible final decay channels 
involving at least one deeply bound molecular state $\overline{FF}'$. Inelastic transitions to this almost continuum of final states 
are allowed for all values of $R$ (see main text). (b) Qualitative representations of the behavior of the hyperradial solutions in the dimer-dimer channel (black dot-dashed curve) 
and in a particular final channel (solid black curve) are shown to illustrate 
that the inelastic transition is more likely near the classical turning point $R_{\lambda}$.}
\label{Potentials}
\end{figure}

The key observation in our model is that the inelastic transitions leading to deeply bound
molecular final states can only occur when {\em at least} three atoms are enclosed at distances comparable to $r_{0}$.
In the hyperspherical representation, the decay pathway
in which only three atoms participate is viewed as an infinite series of avoid crossing between the initial
dimer-dimer channel and all possible final states, as illustrated in Fig.~\ref{Potentials}(a), where the red solid
curve describe the initial collision channel and the green dashed curves some of the possible final states.

Therefore, for a given $a$, the inelastic transition to a particular final state can be described in terms of the Fermi 
Golden rule
\begin{eqnarray}
T_{p}^{(\lambda)}(a)\propto\left\vert \left\langle \Psi_{dd}\left(  R,\Omega
\right)  \left\vert V\left(  R,\Omega\right)  \right\vert \Psi_{\lambda
}\left(  R,\Omega\right)  \right\rangle \right\vert ^{2}
\end{eqnarray}
where $\Psi_{\lambda}$ is the final state wave function, labeled by the quantum number $\lambda$,
$\Psi_{dd}$ is our fully coupled dimer-dimer wavefunction, and $V$ is the sum of the interatomic 
interactions. 
The hyperangular behavior of the integrand is assumed to be proportional to the probability amplitude of three particles being in
close proximity. On the other hand, 
the hyperradial behavior of the outgoing channel will oscillate very quickly away from the classical
turning point, as shown schematically in Fig.~\ref{Potentials}(b). The fast oscillation will, in general, 
cancel out in regions away from the classical turning point, $R_{\lambda}$.
This indicates that the integral over the hyperradius will be proportional to the area within the
first oscillation of $\Psi_{\lambda}\left(  R\right)  $ times the remaining
hyperradial behavior evaluated at $R_{\lambda}$. Based on these considerations, the squared result yields
\begin{eqnarray}
T_{p}^{(\lambda)}(a,R_{\lambda})\propto\dfrac{1}{R_{\lambda}}\left\vert F_{dd}\left(
R_{\lambda}\right)  \right\vert ^{2}{\cal F}\left(R_{\lambda}\right),
\label{TpLambda}
\end{eqnarray}
where $F_{dd}$ is the dimer-dimer hyperradial wave function and ${\cal F}\left(  R_{\lambda}\right)  $ is the 
probability of having three out of four atoms at distances comparable to $r_{0}$  at hyperradius $R_{\lambda}$. 
Therefore, the inelastic transitions to a particular final state $\lambda$ occurs in the vicinity 
of $R_{\lambda}$ and it is governed by the probability of having the three
atoms within distances comparable to $r_{0}$.

In practice, we have calculate ${\cal F}(R_{\lambda})$ by defining the proximity operator 
\begin{eqnarray}
{f}(R,\Omega)=e^{-(r_{12}^2+r_{34}^2)/2r_{0}^2}+~{\rm cyclic~permutaions},
\label{Op}
\end{eqnarray}
which is non-zero only when three atoms are sufficiently close to each other. ${\cal P}$ is then simply defined as
\begin{eqnarray}
{\cal F}(R)\propto \langle\Phi_{dd}(R;\Omega)|{f}(R,\Omega)|\Phi_{dd}(R;\Omega)\rangle,
\end{eqnarray}
where $\Phi_{dd}$ is our fully coupled dimer-dimer channel function and the integration is taken over
all the hyperangles. 

Our model for dimer-dimer relaxation, therefore, is simply obtained by 
summing Eq.~(\ref{TpLambda}) over the near continuum of $\lambda$ states,
approximated by an integral over the classical turning points, which yields 
\begin{eqnarray}
V_{\rm rel}^{dd}&\propto&\frac{1}{k_{dd}}\int T_{p}^{(\lambda)}(a,R_{\lambda}) \rho(R) dR \nonumber\\
&=&\frac{1}{k_{dd}}\int\dfrac{1}{R_{\lambda}}\left\vert F_{dd}\left(
R_{\lambda}\right)  \right\vert ^{2}{\cal F}\left(R_{\lambda}\right) \rho(R_{\lambda}) dR_{\lambda} \nonumber\\
&\approx&\frac{1}{k_{dd}}\int \frac{P_{\rm WKB}(R_{\lambda}){\cal F}(R_{\lambda})}{R_{\lambda} \kappa(R_{\lambda})} \rho(R_\lambda) dR_{\lambda},
\label{Tp}
\end{eqnarray}
where $k^2_{dd}=2m(E+2E_{b})$, $\rho\left(  R_{\lambda}\right)  $ is the nearly constant density of states and $|F_{dd}\left(  R_{\lambda}\right)|^2$ 
was approximated by the WKB wavefunction in the classically forbidden region, 
$|F_{dd}\left(  R_{\lambda}\right)|^2\approx P_{WKB}\left(R_{\lambda}\right)/\kappa(R_{\lambda})$.
Here, $P_{WKB}$ is the WKB tunneling probability for the dimer-dimer hyperradial wavefunction and 
$\kappa^{WKB}\left(R_{\lambda}\right)  $ is the WKB wavenumber.

\begin{figure}[htbp]
\includegraphics[width=3.in,angle=0,clip=true]{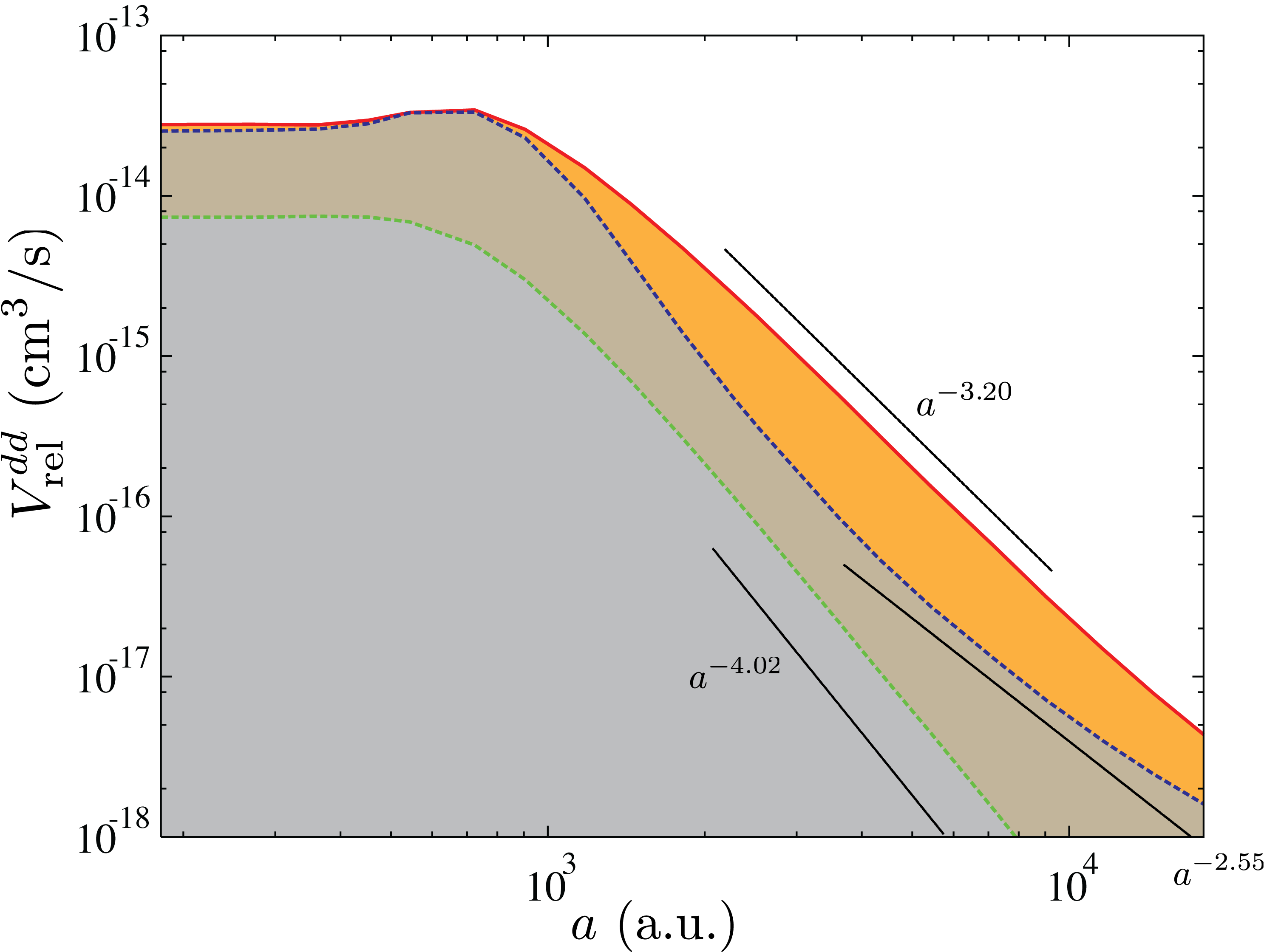}
\caption{(color online). The vibrational relaxation rate for different decay pathways is shown. The red-solid curve is the total rate given by Eq.~(\ref{Tp}), the green-dashed
curve is the contribution from short-range inelastic transitions where all four-atoms participate in the collision. The blue-dashed curve is the contribution
from inelastic transitions near $R=a$ involving only three-atoms illustrating the effects due to the presence of the forth atom (see main text).}
\label{Rates}
\end{figure}

In our model for relaxation, breaking up the integration over $T_{p}(a)$ [Eq.~(\ref{Tp})] 
allows us to analyze the contributions from different pathways. In Fig.~\ref{Rates} we show the total rate as a solid-red curve,
obtained by integrating Eq.~(\ref{Tp}) from 2$r_{0}$ to 10$a$ \cite{MyComment}. In addition to that, we also plot the results for $V_{\rm rel}^{dd}$
obtained by integrating  Eq.~(\ref{Tp}) from 2$r_{0}$ to 5$r_{0}$, see green-dashed curve in Fig.~\ref{Rates}. This result determines
the contribution from inelastic transitions which occur predominantly when all four atoms are within distances comparable to $r_{0}$.
The blue-dashed curve in Fig.~\ref{Rates}, however, shows our results obtained by integrating from $a$ to 4$a$, determining the behavior
of the contributions from inelastic transitions that occurs near $R=a$. The main difference between this contribution and the total rate comes
from the inelastic transitions for $R<a$. Although numerically we are unable to go to larger values of $a$, it is clear that the contributions for 
transitions near $R=a$ becomes increasingly more important and in the very large $a$ limit we expect these contributions to dominate the total rate. 
Interestingly, for the values of $a$ we studied, the contribution for transitions near $R=a$ already falls off slower than the $a^{-2.55}$
prediction of Ref.~\cite{Petrov}. Therefore, we conclude that the mechanism that leads to the $a^{-2.55}$ suppression, although, qualitatively correct,
doesn't quantitatively describe the transitions near $R=a$ due to the presence of the fourth atom.